\documentclass{aa} 
\usepackage{graphicx}  
\def \ha   {H$\alpha$} 
\def \oiii {[O~{\sc iii}]5007~\AA} 
\def \nii  {[N~{\sc ii}]6584~\AA}

\def \tena#1 #2 {\ifmmode{#1 \times 10^{#2}}\else{$#1 \times 10^{#2}$}\fi} 
\def \kms  {\ifmmode{~{\rm km\,s}^{-1}}\else{~km s$^{-1}$}\fi} 
\def \vhel {\ifmmode{~V_{{\rm HEL}}}\else{~$V_{{\rm HEL}}$}\fi} 
\def \vlsr {\ifmmode{~V_{{\rm LSR}}}\else{~$V_{{\rm LSR}}$}\fi} 
\def \vsys {\ifmmode{~V_{{\rm sys}}}\else{~$V_{{\rm sys}}$}\fi} 
\def \vexp {\ifmmode{~V_{{\rm exp}}}\else{~$V_{{\rm exp}}$}\fi} 
 
\def \deg  {\ifmmode{^{\circ}}\else{$^{\circ}$}\fi}  
\def \msun {\ifmmode{{\rm\ M}_\odot}\else{${\rm\ M}_\odot$}\fi} 
\def \myr  {\ifmmode{{\rm\ M}_\odot{\rm\ yr}^{-1}} 
         \else{${\rm\ M}_\odot$ yr$^{-1}$}\fi} 
\def\mnras{MNRAS}

\def\aap{A\&A}

\def\pasp{PASP} 
 
\begin{document} 

\title{A deep mosaic of \oiii\ CCD images of the environment of the LBV star  
P~Cygni}

\author{ P. Boumis\inst{1} 
      \and  J. Meaburn\inst{1,2} 
     \and M. P. Redman\inst{3} 
     \and   F. Mavromatakis\inst{4} 
          } 
 
\offprints{P. Boumis} 
\authorrunning{P. Boumis et al.} 
\titlerunning{P~Cygni} 
\institute{Institute of Astronomy \& Astrophysics, National Observatory of 
Athens, I. Metaxa \& V. Paulou, P. Penteli, GR-15236 Athens, Greece. 
\and Jodrell Bank Observatory, University of Manchester,  
Macclesfield SK11 9DL, UK. 
\and Dublin Institute for Advanced Studies, School of Cosmic Physics,  
5 Merrion Square, Dublin 2, Republic of Ireland. 
\and Technological Education Institute of Crete, General Department of Applied
Science, P.O. Box 1939, GR-710 04 Heraklion, Crete, Greece.
}

\date{Received 17 July 2006 / Accepted 1 August 2006}

\abstract{A mosaic of six, deep, CCD images in the light of the \oiii\ 
nebular emission line has been obtained with the 1.3--m Skinakas (Crete) 
telescope of the filamentary nebulosity surrounding P~Cygni. The 
\oiii\ line discriminates against confusing galactic H{\sc ii} regions 
along the same sight--lines and the new mosaic did not include the 4.8 mag. 
central star; a source of artifacts in the previous lower angular resolution 
observations. New giant `lobes' and `shells' are found to be clustered 
around P~Cygni which must be the relics of historic eruptions  
between 2400 and up to $\approx$~10$^{5}$~yr ago.    
} 
 
\maketitle 
 
\section{Introduction} 
 
The circumstellar environment of the proto-typical Luminous Blue
Variable star (LBV - Conti 1984; Humphreys 1989; Davidson, Moffat \&
Lamers 1989) P~Cygni has been revealed at optical wavelengths in the
work presented in a sequence of seven papers (Johnson et al. 1992;
Barlow et al 1994; Meaburn et al. 1996; O'Connor, Meaburn \& Bryce
1998; Meaburn, L\'{o}pez \& O'Connor 1999; Meaburn et al. 2000;
Meaburn et al. 2004).
  
In these papers the discovery of a 22\arcsec\ diameter inner shell
(IS) was found to be surrounded by a 1.6\arcmin\ diameter, outer shell
(OS). More controversially it was suggested that a large region of
filamentary nebulosity which surrounds the star could have been
ejected in the early stages of its evolution and not simply be an
unrelated supernova remnant along the same sight--line.  The latest
kinematical and morphological evidence in support of this ejection
proposition was presented in Meaburn et al. (2004) as a consequence of
a deep, very wide--field, image in the light of the \oiii\ nebular
emission line, though still some reservations remained about the
association with P~Cygni. The `giant lobe' (GL) projecting to the east
of P~Cygni (O'Connor, Meaburn \& Bryce 1998) was shown to have
southern and eastern counterparts (here these will still be referred
to as GLs). The wide--field CCD \oiii\ image by Meaburn et
al. (2004) was restricted in its sensitivity by the inclusion of the
4.8--mag star P~Cygni in the same field; artifacts due to gross signal
saturation and diffraction became unacceptable in places and areas of
the final image had to be blanked out. Incidentally the
\oiii\ line was chosen for this imagery 
as this is dominant in the emission from the GL nebulosity;
consequently considerable discrimination was achieved against the
predominantly \ha\ and \nii\ emitting, diffuse, galactic H{\sc ii}
regions in the broader field.
 
This saturation/artifact problem has since been overcome by obtaining
a mosaic of deeper \oiii\ CCD images at $\approx$ 5 times higher
angular resolution than previously covered fields adjacent to
P~Cygni but where none includes the star itself. New GL features that
are revealed help to confirm -- and see Sect.3 -- the
association of all of the extensive, filamentary GL nebulosity with
P~Cygni.
 
\section{Observations and results} 
 
A mosaic of 6 images was taken of the adjacent fields to P~Cygni
through an \oiii\ filter with the 1.3 m (f/7.7) Ritchey--Cretien
telescope at Skinakas Observatory in Crete, Greece on September 1--2,
5 and 9--10, 2005. The detector was {\bf a} 1024$\times$1024 (with
24$\times$24 $\mu$m$^{2}$ pixels) SITe CCD giving a field of view of
8\arcmin.5$\times$~8\arcmin.5 and an image scale of 0\arcsec.5
pixel$^{-1}$. One 2400 s exposure in the light of \oiii\ and two 180 s
exposures in the continuum were taken for each field to make sure that
any cosmic ray hits will be identified and removed successfully (see
Boumis et al. 2002 for the image processing details). The
continuum filter was 230 \AA\ wide and centred on 5470 \AA\ and so did
not transmit any strong emission line.
 
The image reduction was carried out using the IRAF package and their
grey-scale representation generated using the STARLINK Kappa and
Figaro packages. The astrometric solutions were calculated for each
field using reference stars from the Hubble Space Telescope (HST)
Guide Star Catalogue (Lasker et al. 1999). All coordinates quoted in
this paper refer to epoch 2000.

Light and dark, negative, grey-scale representations of the same
mosaic of \oiii\ images are shown in Figs. 1 \& 2{\bf ,}
respectively. That in Fig. 1 displays the brighter filaments to
advantage. The continuum image has been subtracted to suppress the
rich star field, whereas in Fig. 2 fainter emission becomes more
apparent. The latter is not continuum subtracted. The central panel of
Fig. 1 containing P~Cygni itself has been filled in with a section of
the original (O'Connor et al. 1998) \nii\ image where an occulting
mask suppressed the bright central star, consequently the eastern, and
newly detected northern (see below) GLs can be seen to be apparantly
projecting from the OS around P~Cygni. The 1.6 arcmin diameter OS
which surrounds P~Cygni is the dark feature in this central panel.
 
All of the \oiii\ emission features shown as contours in fig. 1 of
Meaburn et al. (2004) are confirmed as real here in Figs. 1 \& 2. In
particular, the southern and eastern counterparts to the original GL
are prominent.  However, because of both the higher sensitivity and
resolution, and the absence of artifacts many new and relevant
features are revealed. In Fig. 1 a bright \oiii\ emitting ridge can be
seen projecting from just to the north of P~Cygni. The \nii\ inset
image in Fig. 1 shows that this northern ridge continues south to the
edge of the OS and points directly back to P~Cygni.  Fainter and more
extensive `lobes' $\approx$ 7 \arcmin\ across can now be seen to the
north, east and south of the star. Incidentally, wide absorption lanes
of foreground dust can be also seen in Fig. 1 which cause patchy
extinction of the \oiii\ line over this field with the effect that in
some places only parts of the extensive filamentary features are
observed. Some degree of `bipolar' symmetry exists between pairs of
the faint extensive lobes.  Moreover, the darker greyscale
representation of the same image (now with stars present) in Fig. 2
emphasises how this general \oiii\ emission is concentrated on P~Cygni
unlike any other part of the general field in this emission line.
 
\begin{figure*} 
\centering 
\includegraphics[height=0.6\textheight]{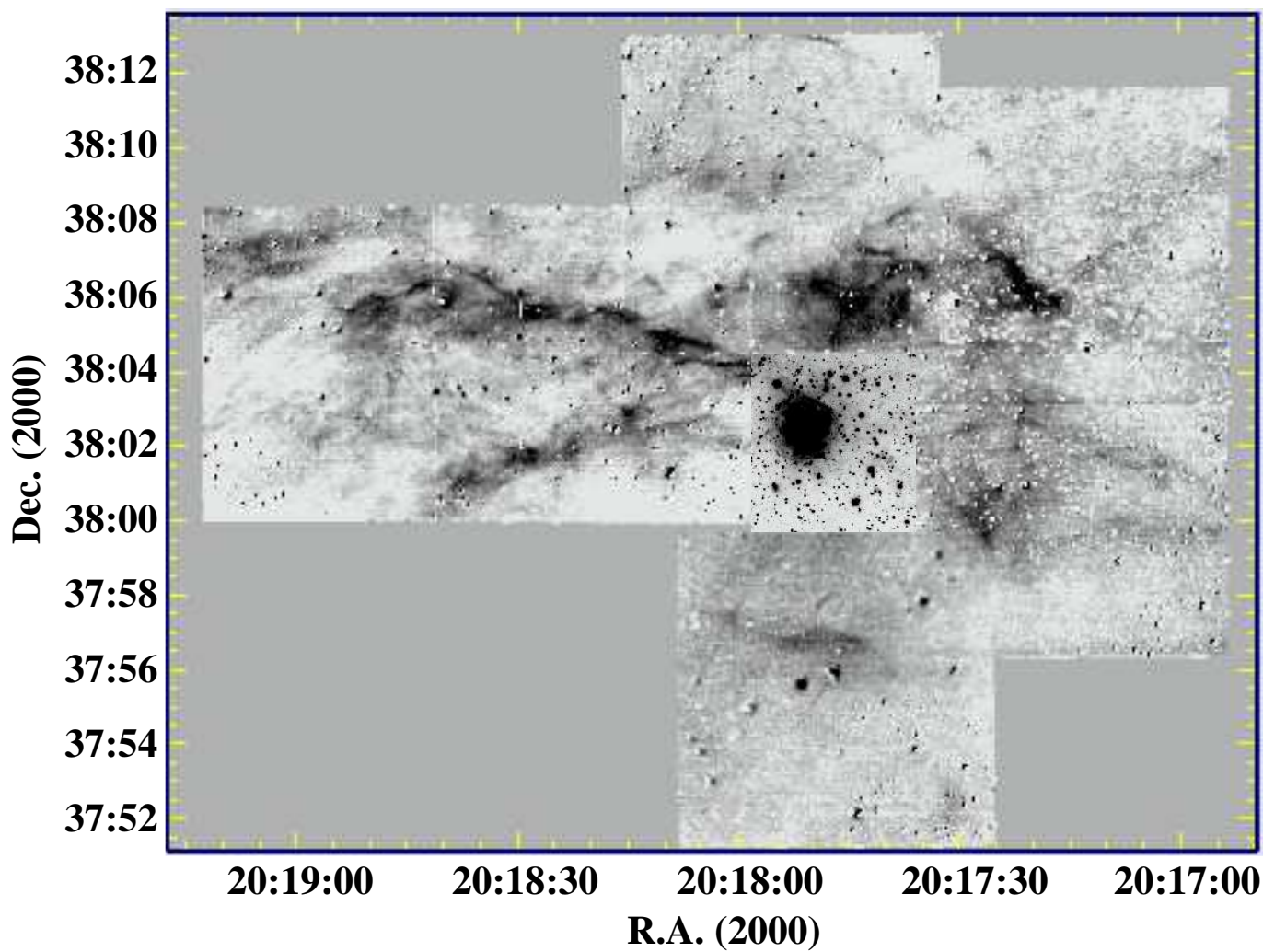} 
\caption{A mosaic of six \oiii\ images of the field around 
P~Cygni is shown. These have been continuum subtracted and only 
residuals from this process, for the brighter stars, are apparent. 
The central panel is filled by part of the original \nii\ image 
from O'Connor et al. (1998) where an occulting mask suppressed 
the image of the 4.8 mag star P~Cygni. This is presented to show the 
extent of the outer shell, OS and the connections to the various 
GLs.} 
\label{Fig1}  
\end{figure*}

\begin{figure*} 
\includegraphics[height=0.6\textheight]{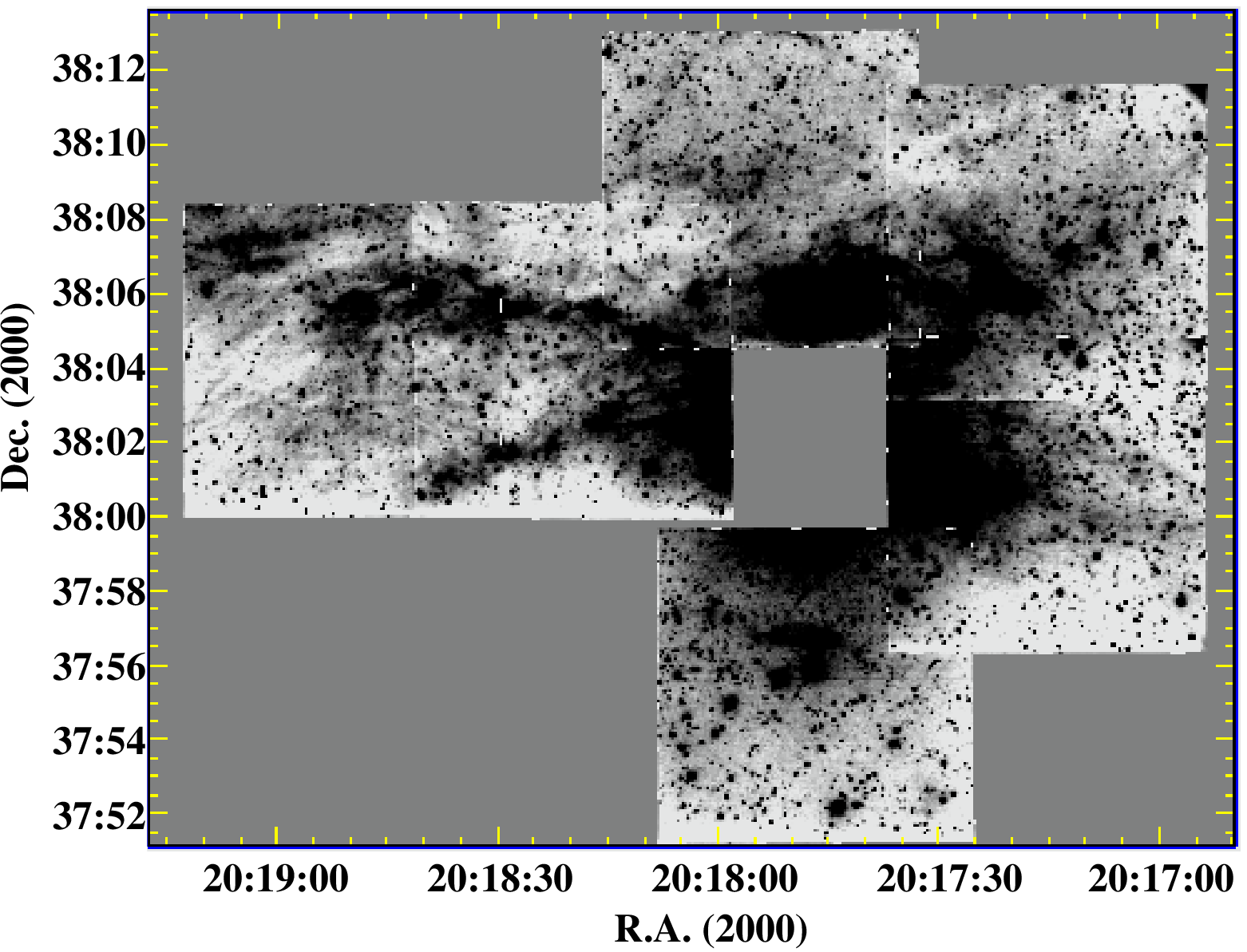} 
\caption{A darker presentation of the same mosaic of \oiii\ that is
shown in Fig. 1 is presented. The continuum light of stars has not
been subtracted and the central panel is no longer filled by the
previous image shown in Fig.1. The concentration of \oiii\ emission
around P~Cygni is emphasised.}  
\label{Fig2}  
\end{figure*} 
 
\section{Discussion} 
 
The wider discussion involving both the morphology and kinematics of
the GL nebulosity around both P~Cygni and related LBV stars in Meaburn
et al. (2004) will not be repeated here. Most of the same conclusions
stand. The purpose of the present paper is to report new observations,
that help to confirm the direct relationship between the surrounding,
extensive, high-excitation filamentary GL nebulosity and P~Cygni.
This association was always left conservatively in doubt after the
previous more limited observations.
 
Here (Figs. 1 \& 2), P~Cygni appears to be completely surrounded by
overlapping, elongated, shell--like filamentary structures over NS and
EW diameters of 16\arcmin\ (8.4 pc) and 20\arcmin\ (10 pc),
respectively, where the distance to this star is 1.8 kpc (van
Schewick 1968; Lamers et al. 1983). The \nii\ and \ha\ line profiles
(Meaburn et al. 1999, 2000 \& 2004) over these GLs have many of the
velocity characteristics of those of the optical filaments of evolved
supernova remnants and, in the limited high spectral resolution
observations so far obtained, exhibit predominantly receding radial
velocities with respect to \vsys\ though coming to this value close to
the star. However, it is this distribution of lobes and shells,
apparently centered on P~Cygni that is compelling evidence for a
direct association. The positive radial velocities found in the
previous limited kinematical observations are still explicable
(Meaburn et al. 2004) as a 'breeze' as P~Cygni moves towards the
observer at high velocity with respect to it local ambient medium.
 
The images in Figs. 1 \& 2 reveal the multiple lobes and shells, each
$\approx$~5pc across, that must be a consequence of continuous
eruptions of P~Cygni over a period of 2400--10$^{5}$ yr ago. The
lower limit is the well--determined (Meaburn et al. 2000) dynamical
age of the OS (shown in the inset in Fig. 1) and the upper limit is
that estimated (Meaburn et al. 2004) for the extremities of the bright
eastern GL also prominant in Fig. 1. All of these features then
precede the observed 1600 AD `great outburst' of P~Cygni (de Groot
1969) but are within the duration of the LBV phase expected for a
50\msun\ star (Humphreys \& Davidson 1994 and note that Turner et
al. 2001 suggested P~Cygni had a progenitor mass as low as
25\msun ). It is now thought most unlikely that the alternative
suggestion, that unrelated SNR optical filaments along the same
sight--line are being observed. It is the symmetry of the \oiii\
shells and lobes, and of the general \oiii\ emission, around P~Cygni
combined with the previous kinematical associations (Meaburn et al
2004) that eliminates this possibility.  Spatially--resolved profiles
of the \oiii\ line are now required over the whole field to
investigate the dynamics of these ejected phenomena more
thoroughly. So far only receding motions have been detected. Also
low--dispersion spectra are required to measure the velocities of the
shocks that are undoubtably exciting these extensive \oiii\ emitting
filaments. Incidentally, the rather unstructured and inhomogeneous
nature of the nebulosity around P~Cygni (compared with planetary
nebulae, for example) is reminiscent of other high--mass evolved
stars; WR 124 is surrounded by a clumpy ring of ejecta that may have
originated in a preceeding LBV phase (Grosdidier et al. 1998;
van der Sluys \& Lamers 2003).

\begin{acknowledgements} 
 
We acknowledge the help of J. Alikakos and S. Akras as well as the
support of the staff at Skinakas observatory during these
observations. Skinakas Observatory is a collaborative project of the
University of Crete, the Foundation for Research and
Technology-Hellas, and the Max-Planck-Institut f\"{u}r
extraterrestrische Physik, Garching. MPR is supported by the
IRCSET, Ireland.
 
\end{acknowledgements}

\end{document}